\documentclass[iop]{emulateapj-rtx4}

\usepackage{txfonts} 
\usepackage{epsfig} 
\usepackage{graphicx}
\usepackage{natbib} 
\usepackage{xspace}
\usepackage{color}
\usepackage[colorlinks,urlcolor=blue,citecolor=blue,linkcolor=blue]{hyperref} 

\newcommand{\dechms}[4]{$#1^{\rm h}#2^{\rm m}#3\mbox{$^{\rm s}\mskip-7.6mu.\,$}#4$}

\newcommand{\decdms}[4]{$+#1^{\circ}#2'#3\mbox{$''\mskip-7.6mu.\,$}#4$}

\newcommand{\be}{\begin{equation}}
\newcommand{\ee}{\end{equation}}

\newcommand{\msun}{{ M_\odot }} 
\newcommand{\kms}{{{\rm km\, s^{-1}}}}  
\newcommand{\yr}{{{\rm yr^{-1}}}} 
\newcommand{\AU}{{{\rm AU}}} 

\begin{document}

\title{Kinematics of the Outflow From The Young Star DG Tau B: \\ Rotation in the vicinities of an optical jet} 

\shortauthors{Zapata, et al.}

\author{Luis A. Zapata\altaffilmark{1}, Susana Lizano\altaffilmark{1}, Luis F. Rodr\'\i guez\altaffilmark{1}, 
Paul T. P. Ho\altaffilmark{2,3}, Laurent Loinard\altaffilmark{1}, \\ Manuel Fern\'andez-L\'opez\altaffilmark{3}, and Daniel Tafoya\altaffilmark{1} }

 \altaffiltext{1}{Centro de Radioastronom\'\i a y Astrof\'\i sica,
UNAM, Apdo. Postal 3-72 (Xangari), 58089 Morelia, Mich., M\'exico}
\altaffiltext{2}{Academia Sinica Institute of Astronomy and Astrophysics, Taipei,
Taiwan}  
\altaffiltext{3}{ Harvard-Smithsonian Center for
Astrophysics, 60 Garden Street, Cambridge, MA 02138, USA}
\altaffiltext{4}{Astronomy Department, University of Illinois, 1002 West Green Street, Urbana, IL 61801, USA}

\begin{abstract} 
We present $^{12}$CO(2-1) line and 1300 $\micron$ continuum observations made with the Submillimeter Array (SMA) 
of the young star DG Tau B.  We find, in the continuum observations, emission arising from the circumstellar
disk surrounding DG Tau B.  The $^{12}$CO(2-1) line observations, on the other hand, revealed emission associated with
the disk and the asymmetric outflow related with this source. Velocity asymmetries about the flow axis are found over 
the entire length of the flow.   The amplitude of the velocity differences is of the order of 1 -- 2 km s$^{-1}$ over distances
 of about 300 -- 400 AU. We interpret them as a result of outflow rotation. The sense of the outflow and disk rotation is 
 the same. Infalling gas from a rotating molecular core cannot explain the observed velocity gradient within 
the flow. Magneto-centrifugal disk winds or photoevaporated disk winds can produce the observed 
rotational speeds if they are ejected from a keplerian disk at radii of several tens of AU.  Nevertheless, 
these slow winds ejected from large radii are not very massive, and cannot account for the observed 
linear momentum and angular momentum rates of the molecular flow. Thus, the observed flow is probably 
entrained material from the parent cloud. DG Tau B is a good laboratory to model in detail the 
entrainment process and see if it can account for the observed angular momentum.
\end{abstract}


\section{Introduction}
Jets and outflows from young stars play a key role in the star-formation process, 
however their nature is still under debate \citep{arce2007}.  
The general consensus is that fast protostellar jets are driven and collimated by rotating magnetic 
fields anchored to the star-disk system \citep{pud2007,sha2007}.  
However, it is not clear where these magnetic fields are anchored to the disk: it could be 
from the radius at which the stellar magnetosphere truncates the disc \citep[{\it X-wind},][]{achu2000} 
or from an extended region of the disc  \citep[{\it disc wind},][]{konigl2000}. 
Magneto-hidrodynamic (MHD) models predict that the  
jet should inherit a toroidal angular momentum component or, in other words, rotation \citep{fen2011}. 
The magnetic field forces the corotation of the gas up to the Alfv\`en radius.
Therefore, the magnitude and position of these movements within the outflow  would give, in principle,
information about their origin on the disk  \citep{and2003}. 
We refer the reader to the introduction of \citet{Zap2010} for details concerning to the previous studies 
on jet/outflow rotation based on optical, infrared and millimeter studies. In this paper \citep{Zap2010}, the results and
limits on previous studies are discussed. 

Located at about 150 pc, DG Tau B is a low-luminosity (0.88 L$_\odot$) 
class I/II young star located about 1$'$ to the SW of the well-known T Tauri star DG Tau 
\citep{jon1986,eis1998,tor2009,luh2010,rod2012}. DG Tau B powers an asymmetric bipolar jet (HH 159) 
mapped at optical/infrared wavelengths  \citep{eis1998}.
The red lobe, located to its NW consists of a chain of bright knots extending to about 1$'$ from the source,
 while the blue lobe located to its SE is much more fainter 
and less collimated,  and is detected only up to  $\sim$10$''$ from the star \citep{mun1983,eis1998,pod2011}.
 A molecular outflow mapped in carbon monoxide (CO) is observed
principally associated with the NW fast optical jet \citep{mit1997}. These authors found that the CO outflow is similar to 
the optical outflow: the CO redshifted emission 
extends at least 6000 AU (40$''$ at the distance of 150 pc) to the NW of the star, while the blueshifted CO 
emission is confined to a compact region, which is less than 500 AU (about 3$''$). \citet{mcg2004}
proposed that the Herbig-Haro objects 836 and 837, located several arcmin to
the SE of DG Tau B, could be tracing ejecta from this star that took place $\sim$10$^4$ years ago.

At the base of the outflow emerging from DG Tau B there is a circumstellar disk placed close to 
edge-on \citep[$\sim$65$^\circ$;][]{gui2011}, and detected for the first time 
in absorption by the broadband imaging of the {\it Hubble Space Telescope} \citep{sta1997}. 
The disk was confirmed by $^{13}$CO millimeter observations \citep{pad1999}. The source position has been determined 
from 3.5 cm VLA radio continuum observations by \citet{rod1995}, who locate it within the 
observed dark lane. The position is at $\alpha_{J2000.0}$ = \dechms{04}{27}{02}{55}, $\delta_{J2000.0}$ = 
\decdms{26}{05}{30}{7}.
The disk has been mapped at millimeter wavelengths using sub-arcsecond resolution 
with the {\it Plateau de Bure Interferometer} (PdBI) \citep{gui2011}. They found a deconvolved size for the 
disk of  $0\rlap.{''}69$ $\times$ $0\rlap.{''}34$ with a P.A. = $26^\circ$, and a mass of 0.068 M$_\odot$.
 \citet{gram2010} modeled the SED of DG Tau B as a young star with a mass of 0.5 M$_\odot$, 
 a temperature of 4000 K and a radius of 2.5 R$_\odot$.

\begin{figure*}[ht]
\begin{center}
\includegraphics[scale=0.45, angle=0]{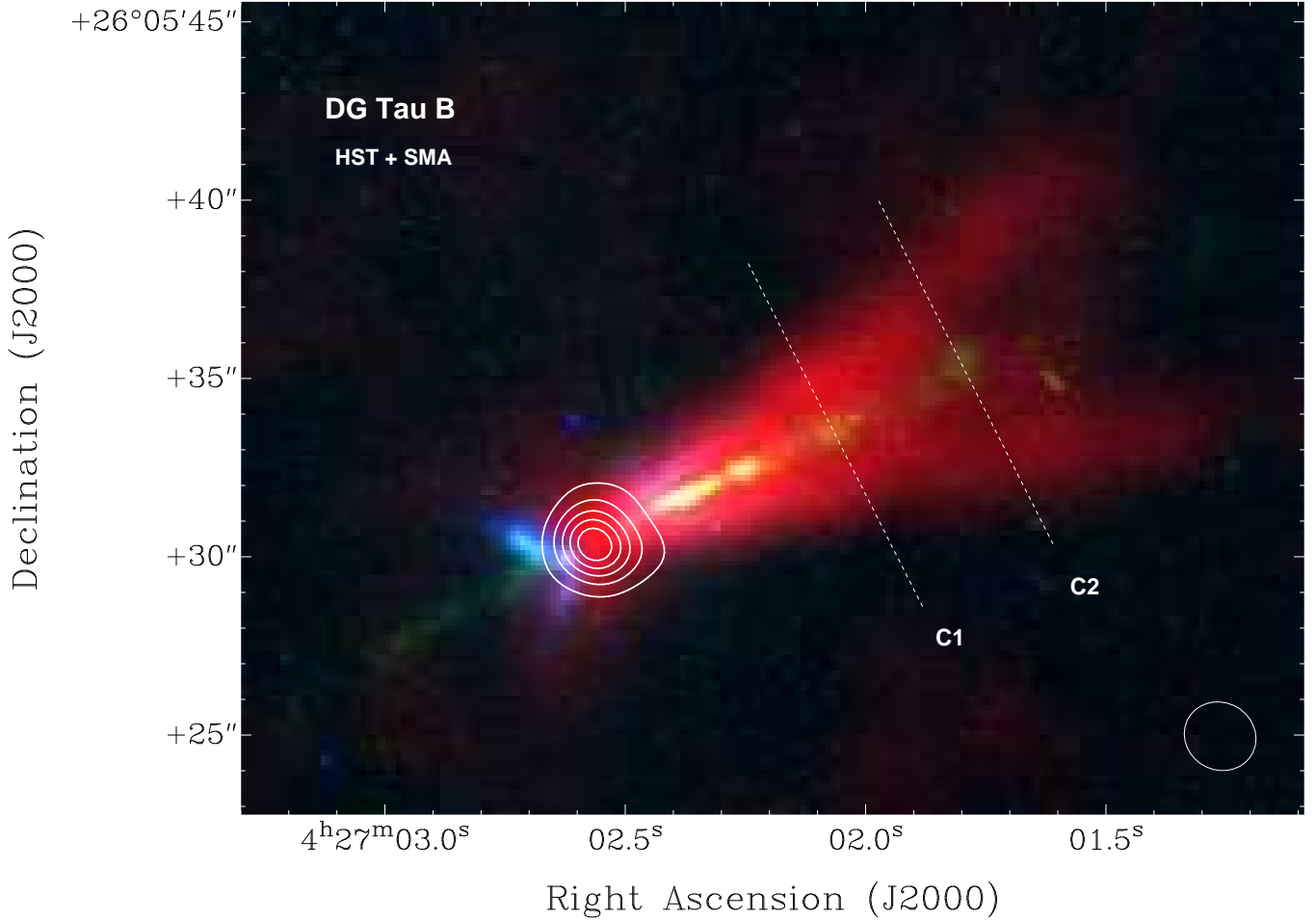}
\caption{ SMA and HST color-coded image of DG Tau B. Color coding: SMA total CO integrated
intensity emission (red),  HST-WFPC2 images with the spectral filters F814w (blue) and F675w (green).
The color-coded image is in addition overlaid with an SMA 1300 $\mu m$ continuum contour image obtained 
by averaging the line-free spectral channels in the upper side band. The velocity range for the CO moment 0 map 
is from $-$2.0 to $+$20.4 km s$^{-1}$.  The white contours range 
from 25\% to 85\% of the peak emission, in steps of 15\%. The peak of the continuum emission 
is 0.31 Jy beam$^{-1}$.  The synthesized beam of the continuum image is shown in the lower right corner.
The two diagonal dashed lines mark the places where the position-velocity diagrams, presented in Figure 2,
were obtained. }
\end{center} 
\label{fig1}
\end{figure*}

In this paper, we present millimeter line and continuum observations, made with the 
{\it Submillimeter Array}\footnote{The Submillimeter Array (SMA) 
is a joint project between the Smithsonian Astrophysical Observatory and the Academia 
Sinica Institute of Astronomy and Astrophysics, and is funded by the Smithsonian Institution and the Academia Sinica.} 
(SMA), of the young star DG Tau B and its associated outflow. These interferometric observations reveal velocity 
asymmetries about the flow axis over the entire length of the flow that can be interpreted 
as outflow rotation. Continuum and line 
observations of the circumstellar
disk associated with DG Tau B are also presented.  The paper is organized as follows:
In \S 2 and 3 we discuss the observations and results. In \S 4 we discuss the 
possible origin of rotating outflows, and in \S 5 we present the conclusions.

\section[]{Observations}

The observations were carried out with the SMA on 2011 December and 2012 February, 
when the array was in its compact and extended configuration, respectively.  The independent baselines in these
configurations ranged in projected length from 10 to 140 k$\lambda$.  The phase reference center 
for the observations was set to $\alpha_{J2000.0}$ = \dechms{04}{27}{02}{66}, $\delta_{J2000.0}$ = 
\decdms{26}{05}{30}{4}. Two frequency bands, centered at 230.457 GHz (Upper Sideband) and 
220.457 GHz (Lower Sideband) were observed simultaneously. We concatenated the two data sets using 
the task in MIRIAD called {\it ``uvcat"}. The two different observations were identical, and only the antenna 
configuration of the SMA changed. The primary beam of the SMA at 
230 GHz has a FWHM of $57''$, and the continuum and line emission arising from DG Tau B
fall well within it.  

The SMA digital correlator was configured to have 48 spectral windows (``chunks'') of 104 MHz and
128 channels each. This provided a spectral resolution of 0.812 MHz ($\sim$ 1 km s$^{-1}$) 
per channel. However, some of these spectral windows included 512 channels (one of these bands included the CO line), providing 
0.203 MHz  ($\sim$ 0.26 km s$^{-1}$) of resolution. This very high spectral resolution allowed 
a reliable study of the CO kinematics.	

Observations of  Uranus provided the absolute scale for the flux density calibration.  
The gain calibrators were the quasars 3C 111 and 3C 84, while 3C 279 was used for bandpass calibration. 
The uncertainty in the flux scale is estimated to be between 15 and 20$\%$, based on the SMA monitoring of 
quasars. 

\begin{figure*}[ht]
\begin{center}
\hspace{-1.0cm}
\includegraphics[scale=0.422, angle=0]{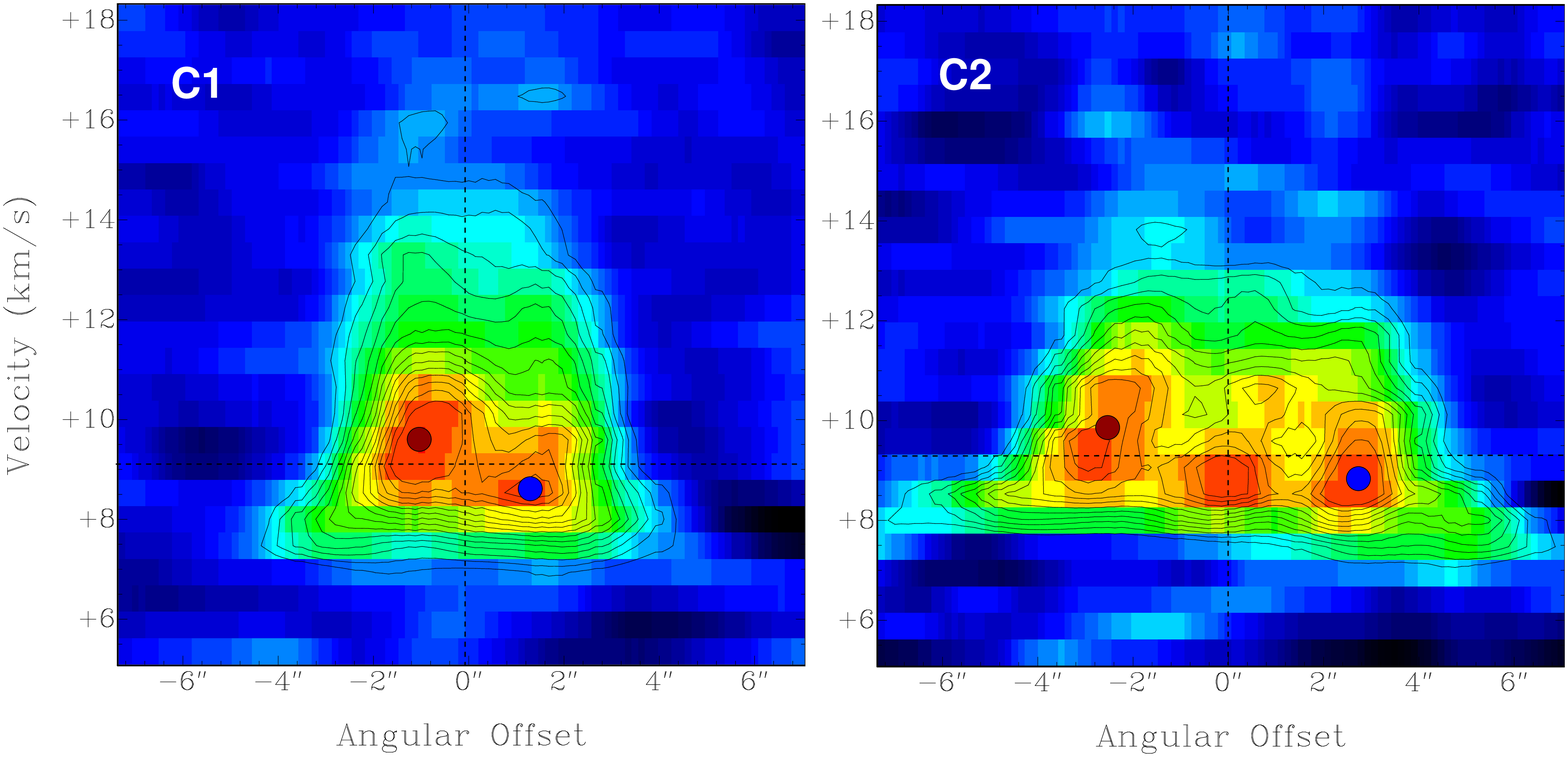}
\includegraphics[scale=0.41, angle=0]{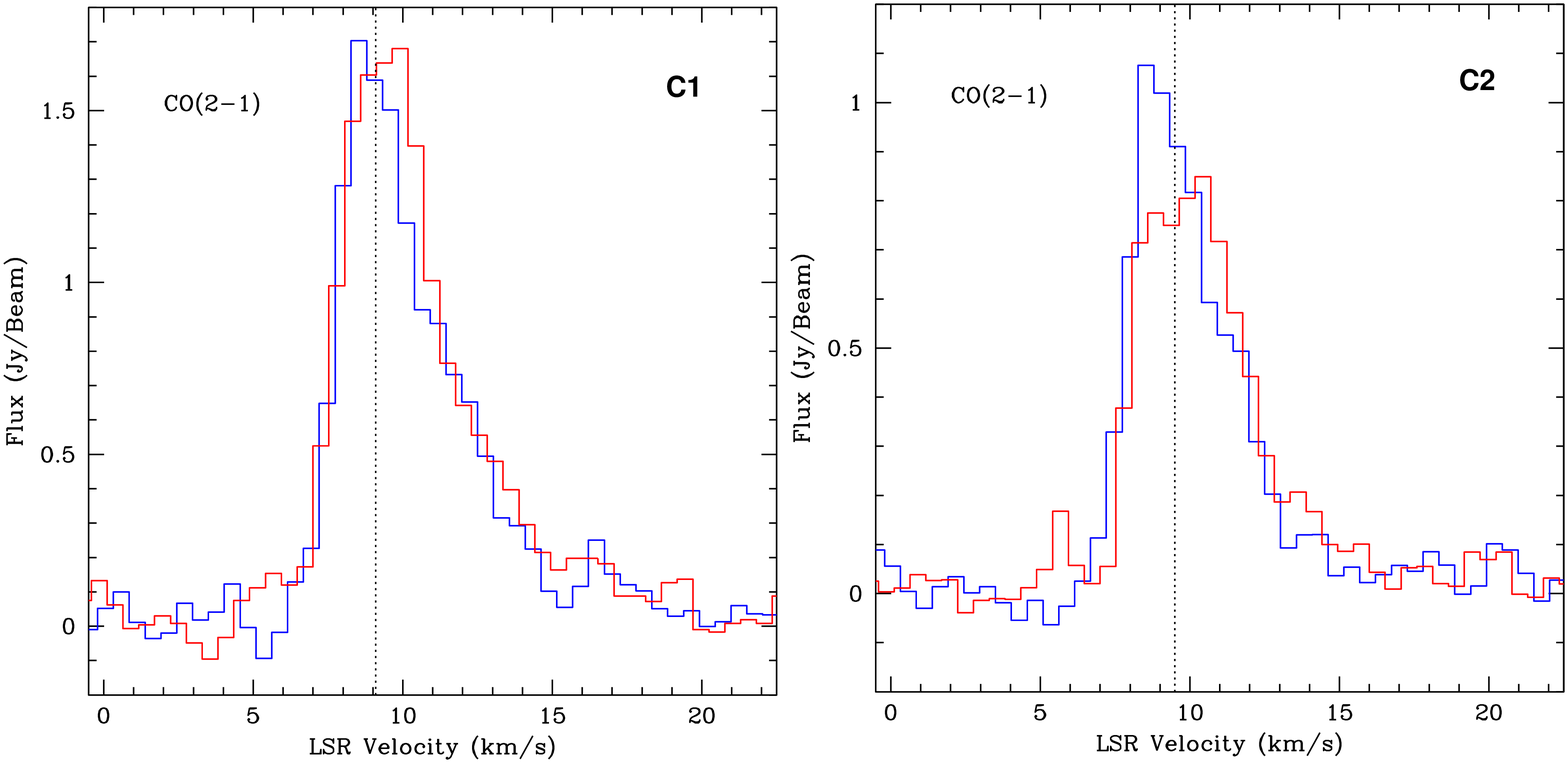}
\caption{{\bf Upper images:} Position-velocity relations across the redshifted side of the molecular outflow in DG Tau B:  
Radial velocity as function of on-the-sky distance. The angular offsets are in arcseconds and the LSR radial velocities 
in km s$^{-1}$. The dashed lines mark the position of the outflow axis and the median velocity of the gradient.   
The molecular gas material in the outflow at these positions has a radial velocity of about $+$9.0 to $+$9.5  km s$^{-1}$. 
The contours in C1 range from 25\% to 90\% of the peak emission, in steps of 5\%. The peak of the line emission 
is 1.9 Jy beam$^{-1}$.  For C2, the contours range from 40\% to 90\% of the peak emission, in steps of 5\%. 
The peak of the emission is 1.05 Jy beam$^{-1}$.  The synthesized beam is 1.87$''$ $\times$ 1.63$''$ with a P.A. 
of $+$64.3$^\circ$, and the spectral resolution is $\sim$ 0.8 km s$^{-1}$. The blue and red dots represent
the positions where the spectra presented in the lower images were obtained. {\bf Lower images:} CO spectra obtained in the positions 
shown in the upper images.  The blue lines correspond to the spectra obtained in the blue dots,
while the red lines to the red dots. The dashed lines mark the position of the median velocity of 
the gradient observed across the outflow. }
\end{center} 
\label{fig2}
\end{figure*}

The data were calibrated using the IDL superset MIR, originally developed for the Owens Valley 
Radio Observatory \citep[OVRO,][]{Scovilleetal1993} and adapted for the SMA.\footnote{The 
MIR-IDL cookbook by C. Qi can be found at http://cfa-www.harvard.edu/$\sim$cqi/mircook.html.} 
The calibrated data were imaged and analyzed in the standard manner using the {\emph MIRIAD} \citep{sau1995} and
{\emph KARMA} \citep{goo96} softwares\footnote{The calibrated data can be obtained from: 
http://www.cfa.harvard.edu/}.  
A 1300 $\micron$ continuum image was obtained by averaging line-free 
channels in the lower sideband with a bandwidth of about 4 GHz. A continuum image using both bands
was also obtained, however similar results were reached.
For the line emission, the continuum was also removed.
We set the {\emph ROBUST} parameter of the task {\emph INVERT} to $+$2 to obtain 
a better sensitivity losing some angular resolution.  The resulting r.m.s.\ noise for the 
continuum was about  9 mJy beam$^{-1}$, 
at an angular resolution of $1\rlap.{''}87$ $\times$ $1\rlap.{''}63$ with a P.A. = $64.3^\circ$. 
The r.m.s.\ noise in each channel of the spectral line data was about 90 mJy beam$^{-1}$ 
at the same angular resolution.

\begin{figure*}[ht]
\begin{center}
\includegraphics[scale=0.44, angle=0]{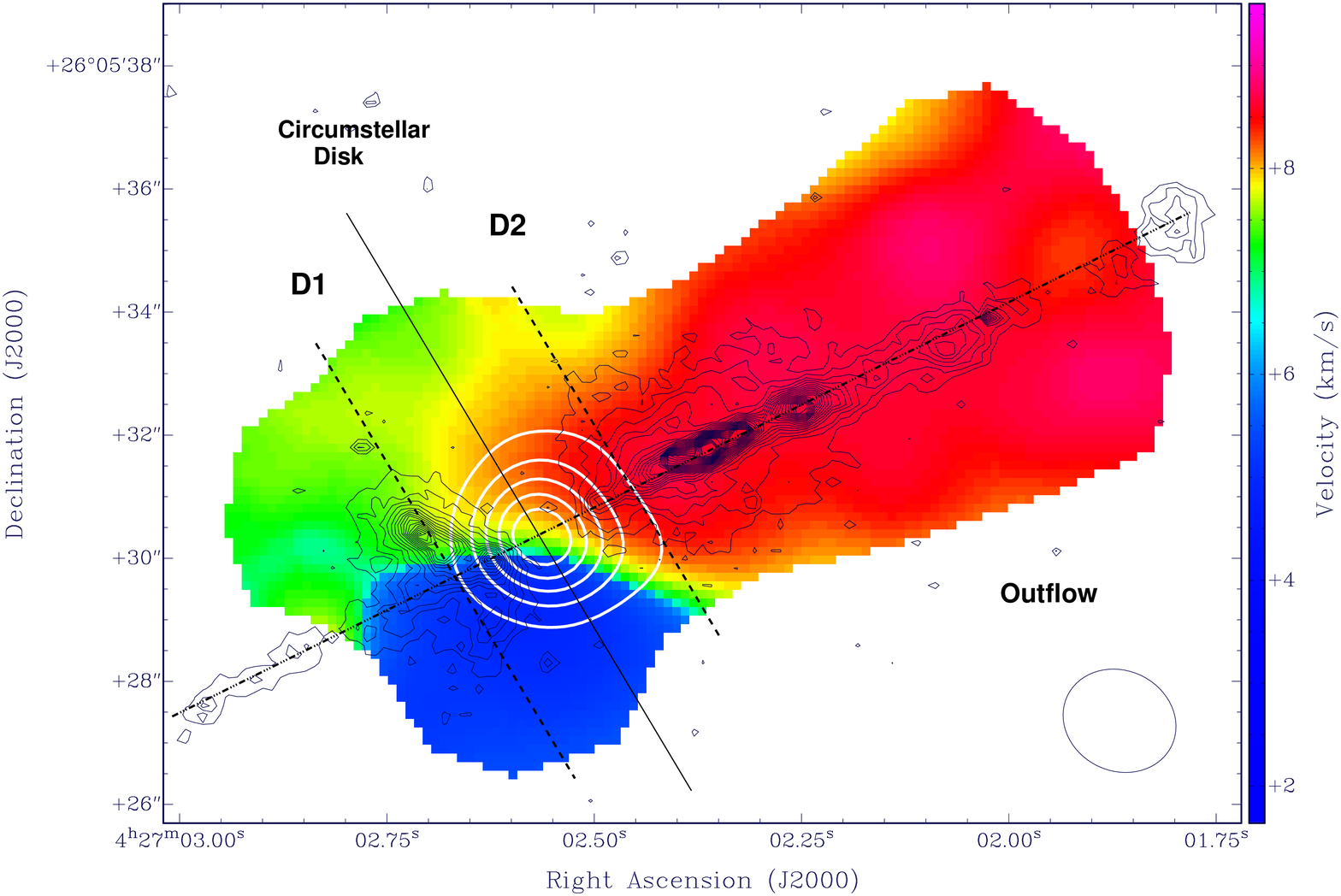}
\includegraphics[scale=0.43, angle=0]{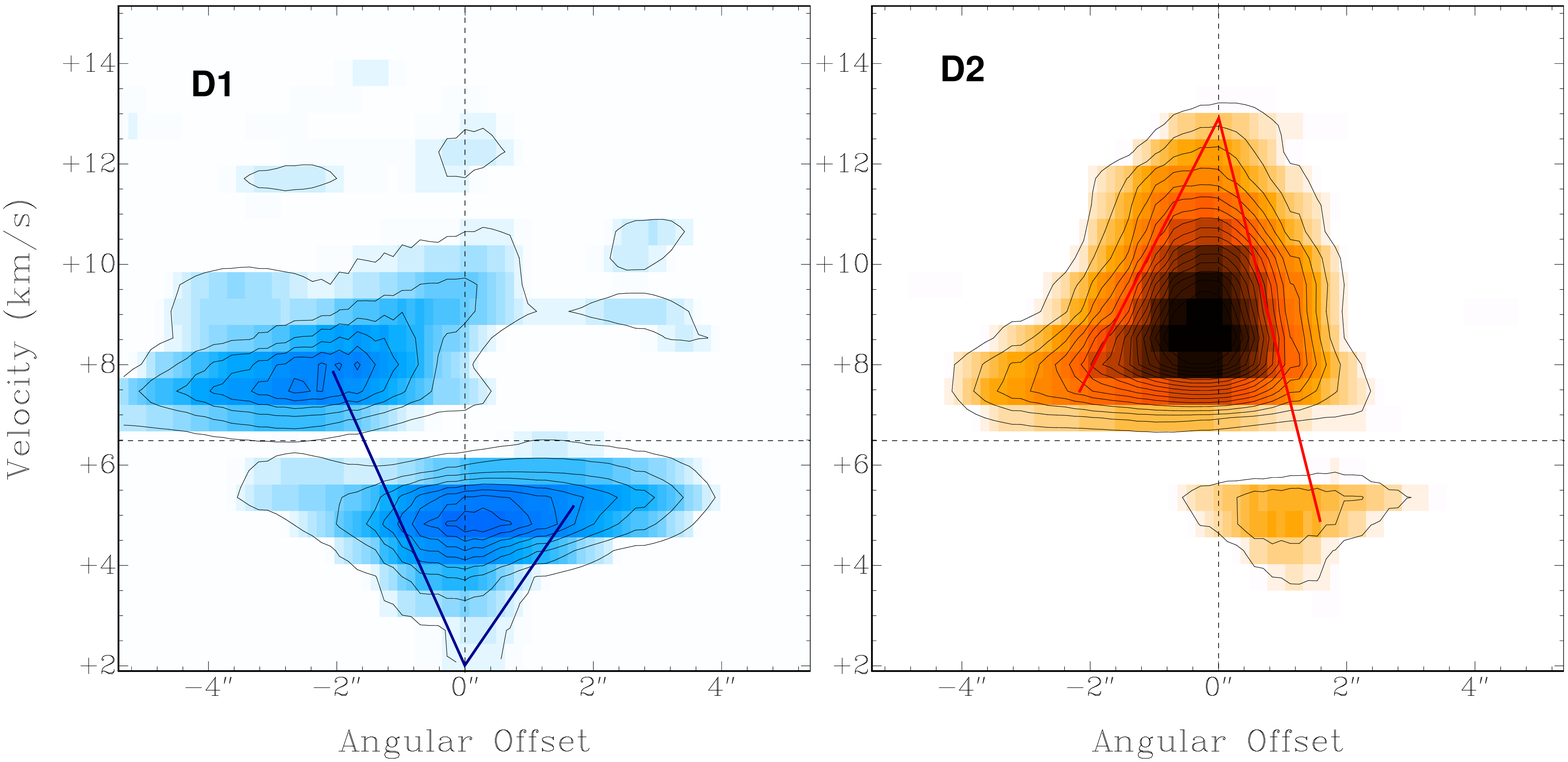}
\caption{{\bf Upper image:} SMA first moment or the intensity weighted velocity of the CO emission from the outflow 
overlaid in black contours with the HST infrared image (using the spectral filter F675w) and in white contours with the 
SMA 1300 $\mu m$ continuum image. The white contours range from 25\% to 85\% of the peak emission, 
in steps of 15\%. The peak of the continuum emission is 0.31 Jy beam$^{-1}$. The synthesized beam of the continuum
image is shown in the lower right corner. The diagonal dashed lines mark the places where the position-velocity 
diagrams, presented here, were made . The color-scale bar on the right indicate the LSR velocities in km s$^{-1}$.
The continuous and dashed lines mark the disk  (P.A. = 208.1$^\circ$) and the outflow (P.A. = 298.1$^\circ$) axis, respectively, 
and the positions of the cuts D1 and D2.
{\bf Lower image:} Position-velocity relations of the CO across the redshifted/blueshifted sides of the molecular 
outflow in DG Tau B, see upper image. The angular offsets are in arcseconds and the radial velocities 
in km s$^{-1}$. The dashed lines mark the position of the outflow axis and the median velocity of the gradient.   
The contours in D1 range from 20\% to 90\% of the peak emission, in steps of 10\%. The peak of the line emission 
is 0.8 Jy beam$^{-1}$.  For D2, the contours range from 40\% to 90\% of the peak emission, in steps of 5\%. 
The peak of the emission is 1.8 Jy beam$^{-1}$.  The synthesized beam is 1.87$''$ $\times$ 1.63$''$ with a 
P.A. of $+$64.3$^\circ$, and the spectral resolution is $\sim$ 0.8 km s$^{-1}$.  The red and blue lines in the lower panels
mark the position of the jet and the velocity gradients.}  
\end{center} 
\label{fig3}
\end{figure*}

\section{Results}

In Figure 1, we present the resulting map of the $^{12}$CO(2-1) line and 1300 $\micron$ continuum 
observations made with the {\it Submillimeter Array} of DG Tau B.  This image is overlaid with {\it Hubble Space Telescope} 
(HST) images obtained with the infrared camera WFPC2 \citep{sta1997}. Figure 1 reveals the innermost parts of the NW 
monopolar redshifted outflow with the rotated ``V" morphology reported for the first time by \citet{mit1997}.  By contrast, the 
blueshifted CO emission is confined to a compact region, close to the disk.  The spatial correspondence between knots in 
the optical jet and successive broadenings of the outflow supports the hypothesis that the molecular flow is produced by 
the action of multiple working surfaces or the entrainment of the ambient gas \citep{mit1997}.  In this image, are also seen the optical jet and the 
infrared nebulosities of scattered light. The 1300 $\micron$ emission falls very well within the  circumstellar disk imaged in absorption 
by the broadband imaging of the HST. We conclude that the 1300 $\micron$ emission here is tracing the innermost parts of the disk, 
very close to the young star DG Tau B.  

For the continuum 1300 $\micron$ emission, and using a Gaussian fitting, we found that the flux density and peak intensity 
values of DG Tau B  are 590$\pm$30 mJy  and 310$\pm$20 mJy beam$^{-1}$, respectively. We also find from this fit 
that the source has a deconvolved size of $1\rlap.{''}8$ $\pm$ $0\rlap.{''}5$ $\times$ $1\rlap.{''}5$ $\pm$ $0\rlap.{''}6$  with 
a P.A. = $+$30$^\circ$ $\pm$ 5$^\circ$. Therefore, at the distance of the Taurus molecular cloud complex (150 pc)
the size of the continuum source is about 270 $\times$ 225 AU.

Following \citet{hil1983} and assuming optically thin isothermal dust emission, a gas-to-dust ratio of 
100, a dust temperature of 30 K, a dust mass opacity $\kappa_{1300 \micron}$ 
= 1.1 cm$^2$ g$^{-1}$ \citep{oss1994},  
and that this object is located at 150 pc, we estimate  
the total mass associated with the dust continuum emission to be 0.08 M$_\odot$, this in a very good agreement 
with the value obtained by \citet{gui2011}. 
The 1300 $\micron$ continuum emission is thus tracing the circumstellar disk surrounding DG Tau B. 
Furthermore, we remark that the mass obtained here could be partially optically thick, and therefore this is a lower limit. 

Following  \citet{zap2014}, assuming local thermodynamic equilibrium (LTE), and that the $^{12}$CO(2-1) molecular emission 
is optically thin, we estimate the outflow mass using the following equation:
$$
\frac{M(H_2)}{M_\odot}=6.3 \times 10^{-20} m(H_2) X_\frac{H_2}{CO}  \left (\frac{c^2  d^2}{2k\nu^2}  \right ) \frac{ \exp \left ( \frac{5.5}{T_{ex}} \right ) \int
S_{\nu} dv \Delta \Omega}{ \left (1 - \exp \left[\frac{-11.0}{T_{ex}} \right ] \right ) }, 
$$ 
where all units are in cgs, m(H$_2$) is mass of the molecular hydrogen, X $_\frac{H_2}{CO}$ 
is the fractional abundance between the carbon monoxide and the molecular hydrogen (10$^4$), 
$c$ is the speed of light, $k$ is the Boltzmann constant, $\nu$ is the rest frequency of the CO line in Hz, $d$ is distance (150 pc), 
$S_{\nu}$ is the flux density of the CO (Jansky), $dv$ is the velocity range in cm s$^{-1}$, 
$ \Delta \Omega$ is the solid angle of the source in steradians,   
and T$_{ex}$ is excitation temperature taken to be 50 K. We then estimate a mass for the outflow powered  
by DG Tau B of 3$\times$10$^{-3}$ M$_\odot$. This value is consistent with the mass of other molecular outflows 
powered by young low-mass protostars, see \citet{wu2004}. The mass estimated here is only a lower limit because 
the CO emission is likely to be optically thick.

In Figure 2, we present two position-velocity diagrams or cuts obtained across the redshifted side of the outflow, 
named C1 and C2, see Figure 1. These diagrams reveal the spatial structure of the gas across the flow as a function 
of the radial velocities and at distances far from DG Tau B ($\sim$ 1400 AU for the case of C1 and 1900 AU for C2).  In both cuts velocity gradients
across the outflow with an amplitude of about 2 km s$^{-1}$ are revealed. In the cut C1, the gradient is somewhat larger than 2 km s$^{-1}$. 
This gradient is also confirmed by the spectra obtained in both sides of the flow as shown in Figure 2 in the lower panels. 
Both cuts were intentionally made over two optical bullets
in order to reveal their dynamics (see Figure 1), however, only the bullet associated with the cut called C2 has a clear molecular
counterpart. This is located in the middle of the borders of the molecular flow.  This molecular emission associated
with the optical bullet is too compact to reveal any velocity gradient.  

In Figure 3 is presented the first moment or the intensity weighted velocity of the $^{12}$CO(2-1) emission from the inner part of the outflow 
overlaid on the HST infrared image (using the spectral filter F675w) and with the SMA 1300 $\mu m$ continuum image.
The image reveals the dynamics of the CO gas in the innermost parts of the flow and in the vicinity of the optical jet. 
Here, the blue colors represent blueshifted emission, while
the red colors, redshifted emission.  The velocity gradient across the outflow is also revealed in this Figure, it is more evident
in its blueshifted side and close to DG Tau B.  A possibility of why the rotation is diluted farther from DG Tau B could come 
from the fact that the flow is not exactly on the plane of the sky, and far from the source the poloidal velocities dominate.       

In Figure 3, in the lower panels, we also show two position-velocity diagrams we made to reveal the kinematics of the molecular gas close to  
the circumstellar disk related with DG Tau B.  In the position-velocity diagrams called D1 and D2 (at a distance from DG Tau B of  about 220 AU), 
are revealed an asymmetric ``V"-shaped 
(or, perhaps, a triangular shape) velocity pattern.  The diagrams are made at a distance of about 250 AU from the source.
These velocity patterns have been attributed to outflow rotation \citep{Lee2008,Zap2010,Pech2012,pet2014}. These patterns are formed because 
the low-velocity gas far from the outflow axis has a velocity gradient ({\it i.e.} rotation), while the gas in the jet and close to the axis seems to be only accelerated in direction of the flow.  If the low-velocity gas were parallel to the direction of the outflow ({\it i.e.} no rotation), one would expect these V shapes to be symmetric with respect to the axis of the flow.  
The amplitude of such velocity asymmetries is about 4 km s$^{-1}$. We noted that this gradient is larger to that observed far from the source,
perhaps due to the conservation of angular momentum.  

From $^{12}$CO (and $^{13}$CO) observations one can
obtain the disk mass and kinematics  (T. Bourke et al. in preparation).
The disk mass is $\sim 0.1 M_\odot$.  Position-velocity
diagrams of both isotopes show disk rotation in
the same sense as the observed outflow rotation. These
diagrams can be fitted by keplerian curves
that give the central stellar mass as a function of
inclination angle. The best fit gives
$M_* \sim 0.5 M_\odot$ for an inclination angle of
65$^\circ$. 

\section{Discussion}

\subsection{Outflow rotation}

For this discussion, the observed velocity gradient across and in the entire length of the molecular outflow from 
DG Tau B is interpreted as rotation.
Even though, as discussed in \citet{Lau2009,Zap2010,Pech2012} there might be some alternative explanations for the observed 
gradient \citep[e.g.][]{sok2005,cer2006}, however, these explanations are unlikely in the case of DG Tau B.  For example, precession or the 
presence of a binary outflow seem to be  ruled out by the existence 
of the collimated and straight optical jet imaged by the HST in the middle of the molecular outflow (see Figure 1).  Also, 
infrared images obtained by \citet{mcg2004} at parsec scales show no evidence of precession for the optical jet. 
The model proposed by \citet[][]{cer2006} is discarded in the case of DG TAU B because the optical jet is not preccesing 
and the velocity gradient is observed in the outflow.

\begin{figure}
\includegraphics[scale=0.43, angle=0]{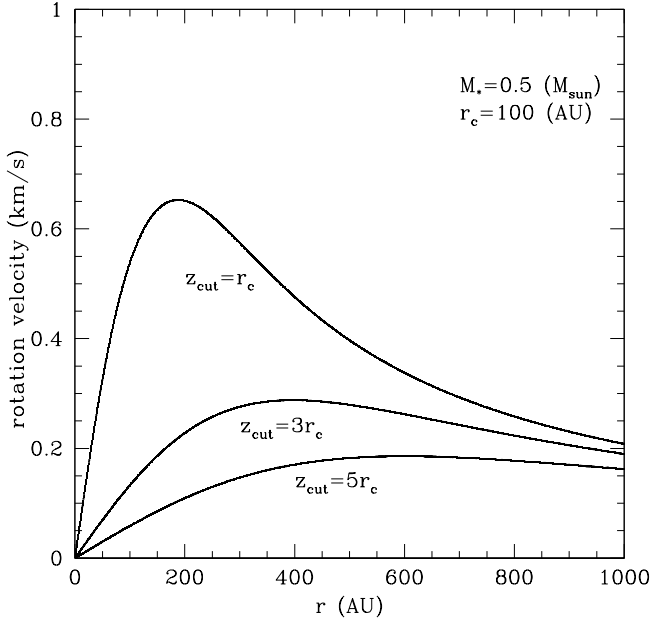}
\caption{Rotation velocity as function of the distance to the rotation axis 
for a rotating collapsing isothermal flow for cuts parallel to the mid plane at different 
heights above the disk, measure in units of the centrifugal radius, $r_c$ = 100 $\AU$.}
\label{fig4}
\end{figure}

\subsection{On the origin of the molecular flow rotation}

From Figures 2 and 3, the observed CO outflow has a  radial velocity of $v_r \sim 3 \, \kms$ (subtracting the LSR velocity of
central source),  and a 
rotation speed $ v_{\varphi} \sim 1 \, \kms $ at a distance from the jet axis  $R \sim 150$ AU. As discussed above,
the outflow rotation is in the same direction as the rotation of the keplerian disk around the central star.
Also, given the inclination angle of the jet axis with respect to the
 l.o.s. $\theta_i \sim 65^o$ \citep{eis1998}, the corrected poloidal speed is 
 $v_p = v_r/\cos(\theta_i)  \sim 7.1 \, \kms$.
 
What is the origin of this rotating molecular flow?
One possibility is that the observed molecular flow is infalling rotating gas from the parent molecular core.
In particular, one can test the collapse flow solution of a rotating isothermal core TSC \citep{Ter1984}.
This flow will collapse to the mid plane and will form  a disk with a size given
by the centrifugal radius, $r_c = \Omega_c^2 G^3 M_*^3 / 16 a^8$, where $\Omega_c$ is the
rotation rate of the core, $G$ is the gravitational constant, $a$ is the sound speed, 
and $M_*$ is the mass of the central star.
Figure \ref{fig4} shows the rotation velocity of the collapsing flow 
as a function of the distance to the rotation axis along cuts parallel to the disk midplane, 
at different heights from the disk: $z=r_c, 3 r_c, 5 r_c$, where we assumed a centrifugal radius $r_c$ = 100 AU.
This figure shows that for all the curves the rotation velocity of the collapsing gas is smaller than the observed rotation
speed $v_\varphi \sim 1 \, \kms$. In fact, for the cut at $z$ = 5 $r_c$ = 500 AU, the rotational speed 
of the infalling gas is  $v_\phi < 0.2 \kms$;
this velocity decreases even more for cuts  further up from the disk midplane. 
Therefore, the observed rotating flow cannot correspond to infalling material from this type of rotating envelope.

Another possibility is that the rotating molecular flow corresponds to a slow wind ejected from the keplerian disk. 
We now examine two types of disk winds: 
magneto-centrifugal winds and photoevaporated winds.
  
For a magneto-centrifugal wind, the poloidal $v_p$ and rotation $v_\phi$  velocities at a distance
$R$ from the rotation axis, can be related to 
the keplerian rotation rate at the footpoint $r_0$ on the disk,
 $\Omega_{K0} = (G M_* / r_0^3)^{1/2}$, by a cubic equation \citep[eq. (4) of 
][]{and2003}. For a central star with a mass $M_*=0.5 \msun$,
this equation gives an footpoint 
$r_0 \sim 11$ AU. 
One can also obtain the parameter $\lambda_\phi = R v_\phi / (r_0^2 \Omega_{K0}) \sim 2.1$,
 which is the ratio of
the specific angular momentum at $R$ to the specific angular momentum at the footpoint $r_0$.
When the jets reach high Alv\'enic Mach numbers and large radii, 
 $\lambda_\phi$  should be equal to the lever arm, $ \lambda = (r_A/r_0)^2$, where
$r_A$ is the Alfv\'en radius where the 
flow reaches the Alfv\'en speed \citep[e.g.,][]{ferreira2006}. 
The lever arm of the flow can be estimated in the asymptotic regime 
from the ratio of the poloidal speed to the 
kepkerian speed at the  footpoint, 
$r_A/r_0 \sim vp/ 2^{1/2} v_K(r_0) \sim 0.8$.  Then,  the implied values of $\lambda_\phi$ and 
 $\lambda$ are very small, with values of $\sim 1 - 2 $, providing very little poloidal 
 acceleration. 
Furthermore, since the lever arm relates the wind mass-loss rate with the disk accretion rate, 
$\dot M_{w} \sim \dot M_{\rm acc}/\lambda$, these low values of $\lambda$ imply a very high efficiency
of mass ejection. In contrast, values of $\lambda \sim 10$ are expected
for MHD disk winds models such that the wind carries away
one tenth of the mass accretion rate \citep[e.g.,][]{konigl2000}. In summary, because the molecular
flow has a  poloidal 
velocity comparable to the rotation speed, its interpretation as a magneto-centrifugal disk
wind poses severe problems of mass ejection.

 The second type of disk winds are photoevaporated winds that arise because the gas at the disk surface is 
 heated by high energy photons from the central star and is able to escape from the local
 gravitational potential well.
Models of these winds from disks around low mass stars
include mainly heating by FUV and X rays
 \citep[e.g.][]{fon2004, gor2009,owen2011}.
These flows have low poloidal speeds $v_p \sim {\rm few} \times a$, where 
the thermal speed is $a \sim  2 \, \kms (T/10^3 K)^{1/2} (\mu/2)^{-1/2}$, where
 $T$ is the gas temperature, and $\mu$  is the mean mass per particle in units of the hydrogen mass $m_H$
 \citep[e.g.,][]{lugo2004}. A characteristic boundary where a photoevaporated wind can escape is given by the
gravitational radius $r_g = G M_*/a^2$, 
although pressure gradients can allow photoevaporation from smaller radii, $r \sim r_g/3 - r_g/2$. 
This radius can be written as
$r_g \sim 108 \left({T/10^3 K } \right)^{-1}  \left({\mu / 2 } \right) \left( { M_* / 0.5 \msun}\right) {\rm AU}.$
Photoevaporated winds are thermal winds and evolve conserving the specific angular momentum 
injected at the disk footpoint, i.e., $ r_0^2 \Omega_{K0}= R v_{\varphi} $. For the observed 
specific angular momentum, this relation gives an ejection point $r_0 \sim 51$ AU  $\sim$  $r_g/2$.


On the other hand, we will now show that magneto-centrifugal and photoevaporated slow disk winds 
do not have enough linear or angular momentum to account for the
observed rates in the molecular outflow.
The observed molecular flow
has a mass $M_{H_2} \sim 10^{-3} \msun$, and the associated dynamical time is 
$t_{\rm dyn} = {1000 \, {\rm AU}  / v_p } \sim 880$ yr. Then, 
the molecular outflow mass-loss rate is $\dot M_{\rm outflow} \sim 1.1 \times 10^{-6} \msun \yr$,
 its  momentum rate is $\dot P_{\rm outflow} =\dot M_{\rm outflow} v_p  \sim  6 \times 10^{-6} \msun \, \yr \, \kms$,
and its angular momentum rate is $\dot L_{\rm outflow} =  \dot M_{\rm outflow} R v_\phi \sim 1.65 \times 10^{-4} \msun\, \AU\, \kms$. 
This linear momentum rate is similar to the rate measured by 
\cite{mitchell1994} and, correcting for partial ionization, 
 is consistent to the momentum rate of the high velocity atomic jet \citep[][]{pod2011}.
Instead, for the slow disk winds considered here, assuming a mass-loss rates similar to that of the
central atomic jet $\dot M_w \sim 10^{-8} \msun\, \yr$, 
 the linear momentum rate is 
$\dot P_w = \dot M_w \, v_p \sim  5.5  \times 10^{-8} \msun \, \yr \, \kms $, and the
 angular momentum rate is 
$\dot L_w = \dot M_{w} r_0 v_{K0} \lambda_\phi = 
\dot M_w r v_\phi = 1.5 \times 10^{-6} \msun\, \AU\, \yr \, \kms$. Both rates
are smaller than
the molecular outflow rates by the factor $ \dot M_{\rm outflow}/M_w \sim 100$.

This large discrepancy between the slow disk wind rates and the outflow rates
indicates that the molecular outflow is probably material from 
a rotating parent cloud entrained by a fast wind (even though the TSC infalling envelope has 
rotational speeds smaller than the 
observed  $v_\phi$).
It would be important to show that such entrainment process is 
capable of carrying also the observed angular momentum \citep[e.g.,][]{canto1996}.
A detailed modeling of this
process is a very interesting problem but is out of the scope of this paper.

\section{Conclusions}

We have reported $^{12}$CO(2-1) line observation of DG Tau B, and discovered velocity asymmetries 
about the flow axis with an amplitude roughly on the order of 2 km s$^{-1}$ far from its exciting source and about 
4 km s$^{-1}$ close to it. This difference in velocities might be due to the conservation of the angular momentum.  
Similar velocity asymmetries are found on both sides of the outflow, and we interpret them as evidence for outflow rotation. 
The outflow rotation sense is identical to that revealed in the circumstellar disk. 
We show that slow disk winds ejected at distances of tens of AU from the star
would have the observed poloidal and rotation velocities. Nevertheless, the observed linear and angular momentum rates
of the outflow are too large compared to the rates of these slow disk winds
and indicate that the molecular rotating flow is probably entrained material form the parent core.
DG Tau B appears to be a promising laboratory for future studies of this entrainment process.

\acknowledgments
L.A.Z, S. L., L. F. R., L.L., and D.T. acknowledge the financial support from DGAPA, UNAM, and CONACyT, M\'exico. 

Facilities: \facility{The Submillimeter Array (SMA)}

\end{document}